\documentclass[onecolumn,showpacs,pra,aps]{revtex4}
\usepackage{dcolumn}
\usepackage{bm}
\usepackage{graphicx}
\usepackage{mathrsfs}

\begin{document}
\title{Distinction of Tripartite Greenberger-Horne-Zeilinger and W States Entangled in Time (or Energy) and Space}
\author{Jianming Wen\footnote{Electronic address: jianm1@umbc.edu\\
Current address: Physics Department, University of Virginia, Charlottesville, Virginia 22904, USA} and Morton H. Rubin}
\affiliation{Physics Department, University of Maryland, Baltimore County, Baltimore, Maryland 21250, USA}

\date{\today}

\begin{abstract}
In tripartite discrete systems, two classes of genuine tripartite
entanglement have been discovered, namely, the
Greenberger-Horne-Zeilinger (GHZ) class and the W class. To date,
much research effort has been concentrated on the polarization
entangled three-photon GHZ and W states. Most studies of
continuous variable multiparticle entanglement have been focused
on Gaussian states. In this Brief Report, we examine two classes
of three-photon entangled states in space and time. One class is a
three-mode three-photon entangled state and the other is a
two-mode triphoton state. These states show behavior similar to
the GHZ and W states when one of the photons is not detected. The
three-mode entangled state resembles a W state, while a two-mode
three-photon state resembles a GHZ state when one of the photons
is traced away. We characterize the distinction between these two
states by comparing the second-order correlation functions
$G^{(2)}$ with the third-order correlation function $G^{(3)}$.
\end{abstract}


\maketitle

\section{Introduction}
Generating entangled states is a primary task for the application
of quantum information processing. The experimental preparation,
manipulation, and detection of multiphoton entangled states is of
great interest for the implementation of quantum communication
schemes quantum cryptographic protocols, and for fundamental tests
of quantum theory. Generation of entangled photon pairs has been
demonstrated from the processes of spontaneous parametric down
conversion (SPDC) \cite{SPDC1,SPDC2,SPDC3} and four-wave mixing
\cite{wen}. These paired photons have proved to be key elements in
many research fields such as quantum computing, quantum imaging,
and quantum lithography. Although entanglement of bipartite
systems is well understood, the characterization of entanglement
for multipartite systems is still under intense study. In
entangled three-qubit states it has been shown that there are two
inequivalent classes of states, under stochastic local operations
and classical communications, namely, the
Greenberger-Horne-Zeilinger (GHZ) class \cite{GHZ} and the W class
\cite{wstate}.

The GHZ class is a three qubit state of the form
$|\mathrm{GHZ}\rangle=\frac{1}{\sqrt{2}}(|000\rangle+|111\rangle)$,
which leads to a conflict between local realism and nonstatistical
predictions of quantum theory. Another three-qubit state, the W
state, takes the form
$|\mathrm{W}\rangle=\frac{1}{\sqrt{3}}(|100\rangle+|010\rangle+|001\rangle)$.
It has been shown that this state is inequivalent to the GHZ state
under stochastic local measurements and classical exchange of
messages \cite{duer}. The entanglement in the W state is robust
against the loss of one qubit, while the GHZ state is reduced to a
product of two qubits. That is, tracing over one of the three
qubits in the GHZ state leaves
$\frac{1}{2}(|00\rangle\langle00|+|11\rangle\langle11|)$, which is
an unentangled mixture state. However, tracing out one qubit in
the W state and the density matrix of the remaining qubits becomes
$\frac{2}{3}|\Psi^{+}\rangle\langle\Psi^{+}|+\frac{1}{3}|00\rangle\langle00|$,
with $|\Psi^{+}\rangle=\frac{1}{\sqrt{2}}(|01\rangle+|10\rangle)$
being a maximally entangled state of two qubits. It has been
further shown that the W state allows for a generalized GHZ-like
argument against the Einstein-Podolsky-Rosen type of elements of
reality \cite{cabello}.

To date, much effort has been concentrated on the polarization
entangled three-photon GHZ and W states. Experimental realizations
of polarization entangled GHZ states and more recently W states
have been performed in optical and trapped ion experiments
\cite{experimentGHZ1,experimentGHZ2,experimentGHZ3,experimentGHZ4,experimentW1,experimentW2,experimentW3}.
Recently, the study of continuous-variable (CV) multipartite
entanglement was initiated in \cite{gaussian2}, where a scheme was
suggested to create pure CV $N$-party entanglement using squeezed
light and $N-1$ beam splitters. In \cite{gaussian} a complete
classification of trimode Gaussian states was with a necessary and
sufficient condition for the separability to determine to which
class a given state belongs. The CV analysis requires
quadrature-type measurement; in this Brief Report we shall be
interested in studying three-photon states using direct photon
counting detection. We here consider three-photon GHZ-type and
W-like states entangled in time and space, which differ from the
CV characterization of \cite{gaussian2}. We will show that
three-mode states, which we denote by $|1,1,1\rangle$, are similar
to W states, while two-mode states, denoted by $|1,2\rangle$,
resemble GHZ-type states. The distinction between these two states
has been demonstrated by looking at the second-order coherence
function $G^{(2)}$.  For related work with emphasis on the
entanglement properties of CV three particle Gaussian GHZ and W
states, see \cite{gaussian3, njp}. This research is of importance,
not only for testing foundations of quantum theory, but also for
many promising applications based on quantum entanglement
\cite{CV,imaging}.

\section{Triphoton W State}
To illustrate the distinction between $|1,1,1\rangle$ and
$|1,2\rangle$ states, we start with the case in which the source
produces three-photon entangled states in different modes. For
simplicity, a monochromatic plane-wave pump beam is assumed to
travel along the $\hat{z}$ direction in the medium producing a state
at the output face of the medium given by
\begin{eqnarray}
|\Psi_1\rangle=\int{d}\omega_1d\omega_2d\omega_3\int{d}\vec{\alpha}_1d\vec{\alpha}_2d\vec{\alpha}_3\Phi(
L\Delta)\delta(\omega_1+\omega_2+\omega_3-\Omega)H(\vec{\alpha}_1+\vec{\alpha}_2+\vec{\alpha}_3)
|1_{\vec{k}_1},1_{\vec{k}_2},1_{\vec{k}_3}\rangle,\label{eq:Wstate}
\end{eqnarray}
where $\Omega$ is the pump frequency, and $\omega_j$ with
$\vec{\alpha}_j$ are the frequencies and transverse wave vectors of
photons in mode $\vec{k}_j$, respectively.
$\delta(\omega_1+\omega_2+\omega_3-\Omega)$ is the steady-state or
the frequency phase-matching condition. The integral over the finite
length $L$ of the system gives the longitudinal detuning function,
$\Phi(L\Delta)$, which determines the natural spectral width of the
triphoton state. The longitudinal detuning function, in the
non-depleted pump approximation usually takes the form of
\begin{eqnarray}
\Phi(x)=\frac{1-e^{-ix}}{ix}=\mathrm{sinc}\Big(\frac{x}{2}\Big)e^{-i(x/2)},\label{eq:sinc1}
\end{eqnarray}
with  $x=L\Delta$ and
$\Delta=(\vec{k}_p-\vec{k}_1-\vec{k}_2-\vec{k}_3)\cdot\hat{z}$,
and $\vec{k}_p$ is the wave vector of the input pump field. Let
$\omega_j=\Omega_j+\nu_j$ with fixed frequency $\Omega_j$.
Choosing the central frequencies so that
$\Omega=\Omega_{1}+\Omega_{2}+\Omega_{3}$, frequency phase
matching now becomes $\nu_1+\nu_{2}+\nu_{3}=0$. Assuming
$|\nu_j|<<\Omega_j$, and that the crystal is cut for collinear
phase matching, $k_{p}=K_{1}+K_{2}+K_{3}$, we can expand $k_{j}$
in powers of $\nu_{j}$,  $k_{j}=K_{j}+\nu_{j}/u_{j}+\cdots$ where
$1/u_{j}$ is the group velocity of the photon $j$ evaluated at
$\Omega_{j}$. Then to leading order we may write $x$ as
\begin{equation}
x=-\sum_{j=1}^{3}L\nu_{j}/u_{j}=-\nu_1L/D_{12}-\nu_3L/D_{32}, \label{eq:DeltaL}
\end{equation}
where we have used frequency phase matching to eliminate
$\nu_{2}$, and  $1/D_{ij}$ is the time difference between the
$i$th photon and the $j$th one passing through a unit length
material. With a slight abuse of notation, we shall write
$\Phi(L\Delta)=\Phi(\nu_1,\nu_3)$. The integration over the
transverse coordinates ($\vec{\rho}$) on the output surface(s) of
the source gives the transverse detuning function as
\begin{eqnarray}
H(\vec{\alpha}_1+\vec{\alpha}_2+\vec{\alpha}_3)=\frac{1}{A}\int{d}\vec{\rho}e^{i\vec{\rho}\cdot(\vec{\alpha}_1+\vec{
\alpha}_2+\vec{\alpha}_3)}.\label{eq:trans}
\end{eqnarray}
In the ideal case, $H$ becomes a $\delta$-function,
$\delta(\vec{\alpha}_1+\vec{\alpha}_2+\vec{\alpha}_3)$. In
Eq.~(\ref{eq:Wstate})  we use the paraxial approximation, which is
a good approximation for quantum imaging and lithography
\cite{rubin,wen12}.  With the quasi-monochromatic assumption
$|\nu_{j}|<<\Omega_{j}$ this leads to the factoring of the state
into longitudinal and transverse degrees of freedom in the
quasi-monochromatic approximation. We are interested in examining
the temporal and spatial correlations between two subsystems by
tracing the third in the free-propagation geometry. The
second-order [$G^{(2)}$] and third-order [$G^{(3)}$] correlation
functions are defined, respectively, as
\begin{eqnarray}
G^{(2)}&=&\sum_{\vec{k}_3}|\langle0|a_{\vec{k}_3}E^{(+)}_2E^{(+)}_1|\Psi_1\rangle|^2,\label{eq:correlation}\\
G^{(3)}&=&|\langle0|E^{(+)}_3E^{(+)}_2E^{(+)}_1|\Psi_1\rangle|^2,\label{eq:correlation3}
\end{eqnarray}
with freely propagating electric fields given by
\begin{eqnarray}
E^{(+)}_j(\vec{\rho}_j,z_j,t_j)=\int{d}\omega_j\int{d}\vec{\alpha}_jE_jf_j(\omega_j)e^{-i\omega_jt_j}e^{i(k_jz_j+\vec{
\alpha}_j\cdot\vec{\rho}_j)}a_{\vec{k}_j},\label{eq:electricfield}
\end{eqnarray}
where $E_j=\sqrt{\hbar\omega_j/2\epsilon_0}$, $k_j=\omega_j/c$ is
the wave number, $z_j$ and $\vec{\rho}_j$ are spatial coordinates of
the $j$th detector, and $a_{\vec{k}_j}$ is a photon annihilation
operator at the output surface of the source and obeys
$[a_{\vec{k}},a^{\dagger}_{\vec{k}'}]=\delta(\vec{\alpha}-\vec{\alpha}')\delta(\omega-\omega')$,
respectively. The function $f_j(\omega)$ is a narrow bandwidth
filter function which is assumed to be peaked at $\Omega_j$. In
Eq.~(\ref{eq:electricfield}) we have decomposed $\vec{k}_j$ into
$k_j\hat{z}+\vec{\alpha}_j$.

Substituting Eqs.~(\ref{eq:Wstate}) and (\ref{eq:electricfield})
into (\ref{eq:correlation}) gives
\begin{eqnarray}
G^{(2)}=C_0G^{(2)}_l(\tau_1-\tau_2)\times{G}^{(2)}_t(\vec{\rho}_1-\vec{\rho}_2),\label{eq:G2}
\end{eqnarray}
where $C_0$ is a slowly varying constant, and the temporal and
spatial correlations, respectively, are
\begin{eqnarray}
G^{(2)}_l(\tau_1-\tau_2)&=&\int{d}\nu_3\bigg|\int{d}\nu_1f_1(\nu_1)f_2(\nu_1+\nu_3)\Phi(\nu_1,\nu_3)e^{-i\nu_1(\tau_1-
\tau_2)}\bigg|^2,\label{eq:temporal}\\
G^{(2)}_t(\vec{\rho}_1-\vec{\rho}_2)&=&\int{d}\vec{\alpha}_3\bigg|\int{d}\vec{\alpha}_1e^{i\vec{\alpha}_1\cdot(\vec{\rho}
_1-\vec{\rho}_2)}\bigg|^2,\label{eq:transverse}
\end{eqnarray}
where $\tau_j=t_j-z_j/c$ and $\omega_j=\Omega_j+\nu_j$. Similarly,
plugging Eqs.~(\ref{eq:Wstate}) and (\ref{eq:electricfield}) into
(\ref{eq:correlation3}) yields
\begin{eqnarray}
G^{(3)}=C_1G^{(3)}_l(\tau_1-\tau_2,\tau_3-\tau_2)\times{G}^{(3)}_t(\vec{\rho}_1-\vec{\rho}_2,\vec{\rho}_3-\vec{\rho}_2),
\label{eq:G3}
\end{eqnarray}
where the third-order temporal and spatial correlations are
\begin{eqnarray}
G^{(3)}_l(\tau_1-\tau_2,\tau_3-\tau_2)&=&\bigg|\int{d}\nu_1d\nu_3f_1(\nu_1)f_2(\nu_1+\nu_3)f_3(\nu_3)\Phi(\nu_1,\nu
_3)e^{-i\nu_1(\tau_1-\tau_2)}e^{-i\nu_3(\tau_3-\tau_2)}\bigg|^2,\label{eq:temporal3}\\
G^{(3)}_t(\vec{\rho}_1-\vec{\rho}_2,\vec{\rho}_3-\vec{\rho}_2)&=&\bigg|\int{d}\vec{\alpha}_1d\vec{\alpha}_3e^{i
\vec{\alpha}_1\cdot(\vec{\rho}_1-\vec{\rho}_2)}e^{i\vec{\alpha}_3\cdot(\vec{\rho}_3-\vec{\rho}_2)}\bigg|^2,
\label{eq:transverse3}
\end{eqnarray}
and $C_1$ is constant. By comparing Eq.~(\ref{eq:temporal}) with
(\ref{eq:temporal3}), it is clear that although one photon is not
detected (traced away) in the two-photon detection, there remains
a correlation between the remaining two photons. The width of the
two-photon temporal correlation depends on the three photon
bandwidth. The comparison between Eqs.~(\ref{eq:transverse}) and
(\ref{eq:transverse3}) indicates that the spatial correlation
between two photons is limited by the bandwidth of the transverse
modes. Ideally, point-to-point correlation is achieved by assuming
infinite transverse bandwidth. Combining the temporal and spatial
properties together show that the $|1,1,1\rangle$ state
(\ref{eq:Wstate}) is a W state entangled in time and space, which
is robust against one-photon loss.
\begin{figure}[tbp]
\includegraphics[scale=0.7]{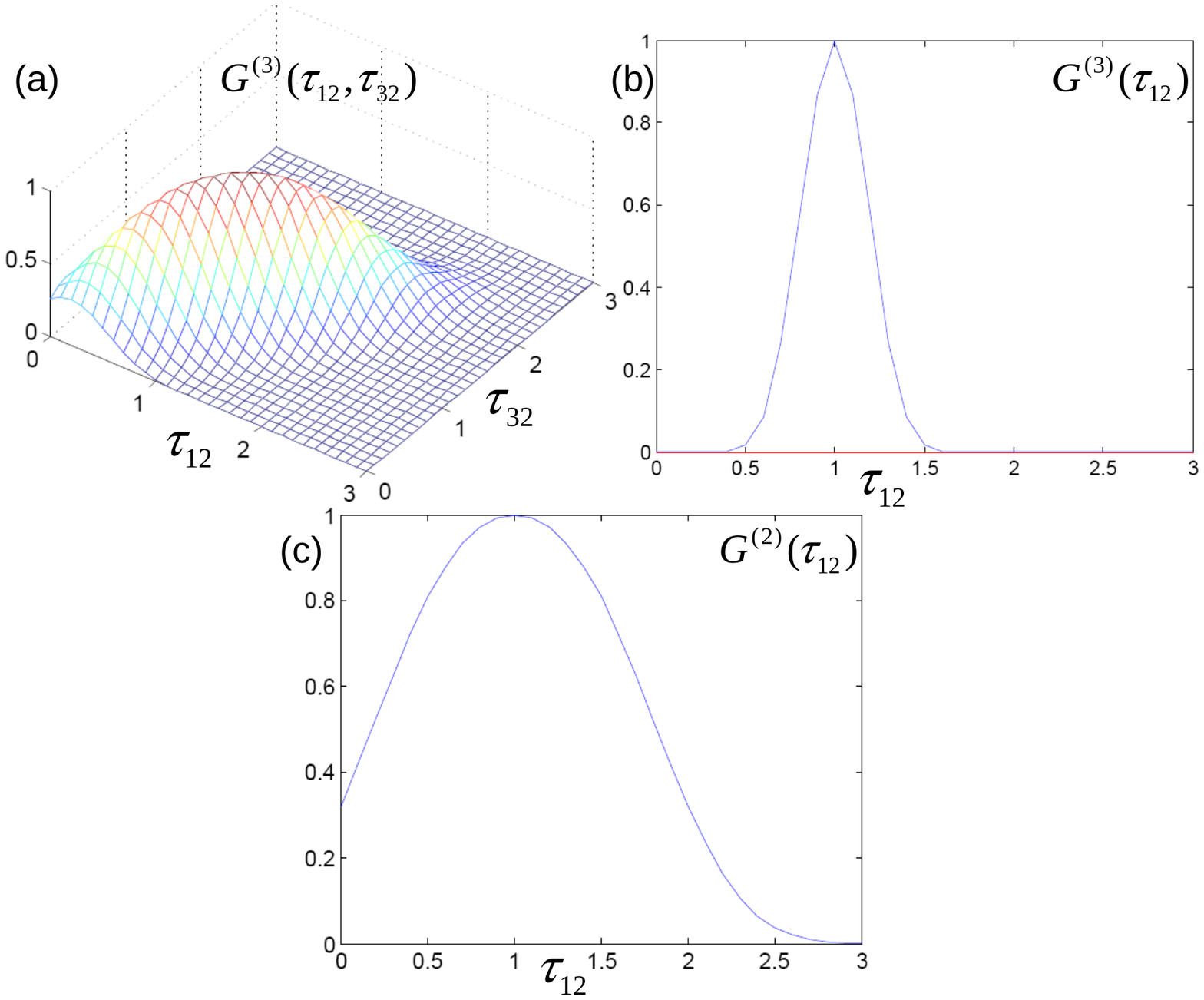}
\caption{(color online) Temporal correlations of $G^{(3)}_l$ and
$G^{(2)}_l$ for the $|1,1,1\rangle$ state normalized to unity at
their origin. The units of  $\tau_{ij}=\tau_i-\tau_j$ are 10 ps. (a)
Third-order temporal correlation $G^{(3)}_l(\tau_{12},\tau_{32})$.
(b) Conditional third-order correlation $G^{(3)}_l(\tau_{12})$
obtained by setting $\tau_{32}=-\tau_{12}+|L/D_{12}|$. (c)
Second-order temporal correlation $G^{(2)}_l(\tau_{12})$.  The
corresponding parameters are chosen as $L/2D_{ij}=10$ ps and all the
filters are Gaussian with the same bandwidth of 0.4
THZ.}\label{fig:GHZ}
\end{figure}

There are several schemes which might produce such a state. One
scheme is three-photon cascade emission whose spectral properties
have been analyzed in \cite{chekhova}. Another configuration
utilizes two parametric down conversions and one up-conversion to
create a triphoton state, as proposed by Keller \textit{et al}
\cite{keller}. The transverse properties of triphotons generated
from such a case have been studied in \cite{wen1} by considering
quantum imaging experiments. It was shown that by implementing
two-photon imaging, the quality of the images is limited by the
bandwidth of the transverse modes of the non-detected third photon.

In Fig.~1 we have compared the temporal correlations between the
third-order correlation function $G^{(3)}_l(\tau_{12},\tau_{32})$
and the second-order $G^{(2)}_l(\tau_{12})$ with Gaussian filters
in Eqs.~(\ref{eq:temporal}) and (\ref{eq:temporal3}). The filters
were taken to the same bandwidth which is large compared to the
width of the $\Phi(L\Delta)$ function. The plots have been
normalized with respect to their  maximum value. In generating the
figure $D_{12}$ has been taken equal to $D_{32}$ and they have
both been taken to be negative.  Because of this the plot of
$G^{(3)}$ is symmetric around the line $\tau_{12}=\tau_{32}$, and
only positive values of the $\tau_{ij}$ are physically allowed.
The length of $G^{(3)}$ is determined by the phase matching
function $\Phi$ as illustrated in Fig.~1(a); Fig.~1(b) shows the
conditional measurement of $G^{(3)}_l(\tau_{12})$ obtained by
setting $\tau_{32}=-\tau_{12}+|L/D_{12}|$. The width of
$G^{(3)}_l$ is determined by the filters. In Fig.~1(c) the
second-order temporal correlation $G^{(2)}_l(\tau_{12})$ is
plotted.  The width of $G^{(2)}_l(\tau_{12})$ is larger than that
of the conditional $G^{(3)}_l(\tau_{12})$ reflecting the lack of
cutoff of the bandwidth for the non-detected third photon.

\section{Triphoton GHZ State}
After analyzing the properties of the $|1,1,1\rangle$ state, we now
consider the case in which the source produces three-photon
entangled states with a pair of degenerate photons of the form
\cite{wen2}
\begin{eqnarray}
|\Psi_2\rangle=\int{d}\omega_1d\omega_2\int{d}\vec{\alpha}_1d\vec{\alpha}_2\Phi(x)\delta(2\omega_1+
\omega_2-\Omega)\delta(2\vec{\alpha}_1+\vec{\alpha}_2)|2_{\vec{k}_1},1_{\vec{k}_2}\rangle,\label{eq:GHZstate}
\end{eqnarray}
where $\Phi$ characterizes the natural bandwidth of triphotons and
has the same form as Eq.~(\ref{eq:sinc1}), with
$x=-2L\nu_1/D_{12}$, $\omega_{j}=\Omega_j+\nu_j$, and
$\vec{\alpha}_{j}$ are the frequencies and transverse wave vectors
of the degenerate $(j=1)$ and nondegenerate $(j=2)$ photons. In
\cite{wen2} we show that by sending two degenerate photons to the
target while keeping the non-degenerate one traversing the imaging
lens, a factor-of-2 spatial resolution improvement can be
obtained, beyond the Rayleigh diffraction limit. Before proceeding
with the discussion, we note that the major difference between the
$|1,2\rangle$ state [Eq.~(\ref{eq:GHZstate})] and $|1,1,1\rangle$
[Eq.~(\ref{eq:Wstate})] is that the $|1,1,1\rangle$ state has more
degrees of freedom than the $|1,2\rangle$ state. This is the
source of the difference between two states when performing
two-photon detection, as we shall see. Physically, because two of
the photons are degenerate, the measurement of one of them
separately uniquely determines the state of the other one and the
two photon state becomes a product state. This is true even if the
photon is not measured but can be measured separately in
principle.   The effect of this is that the state generated is a
mixed state.  Note that for the completely degenerate case, a
similar argument implies that tracing away one of the photons
gives a mixed two-photon state.

For the two-photon measurement here, we first assume that one of the
degenerate photons is not detected. The second-order $G^{(2)}$ and
third-order $G^{(3)}$ correlation functions now become
\begin{eqnarray}
G^{(2)}&=&\sum_{\vec{k}_1}|\langle0|a_{\vec{k}_1}E^{(+)}_2E^{(+)}_1|\Psi_2\rangle|^2,\label{eq:correlation2}\\
G^{(3)}&=&|\langle0|E^{(+)}_2[E^{(+)}_1]^2|\Psi_2\rangle|^2,\label{eq:correlation4}
\end{eqnarray}
where $E^{(+)}_j$ is the free-space electric field given in
Eq.~(\ref{eq:electricfield}). Note that because of the degeneracy, a
two-photon detector is necessary for three-photon joint detection
\cite{wen2}. Following the same procedure for the $|1,1,1\rangle$
calculation, it is easy to show that the second-order and
third-order correlation functions are
\begin{eqnarray}
G^{(2)}_l&=&\int{d}\nu_1\bigg|f_1(\nu_1)f_2(\nu_1)\Phi(-2\nu_1/D_{12})\bigg|^2,\label{eq:temporal2}\\
G^{(3)}_l(\tau_{12})&=&\bigg|\int{d}\nu_1f^2_1(\nu_1)f_2(\nu_1)\Phi(-2\nu_1/D_{12})e^{-2i\nu_1\tau_{12}}\bigg|^2,
\label{eq:temporal4}
\end{eqnarray}
in the temporal domain, and
\begin{eqnarray}
G^{(2)}_t&=&\int{d}\vec{\alpha}_1,\label{eq:transverse2}\\
G^{(3)}_t(\vec{\rho}_1-\vec{\rho}_2)&=&\bigg|\int{d}\vec{\alpha}_1e^{2i\vec{\alpha}_1\cdot(\vec{\rho}_1-\vec{\rho}_2)}
\bigg|^2,\label{eq:transverse4}
\end{eqnarray}
in the spatial space. Comparing Eqs.~(\ref{eq:temporal2}) and
(\ref{eq:transverse2}) with (\ref{eq:temporal4}) and
(\ref{eq:transverse4}) shows that if one of the degenerate photons
is traced away, there will be no correlation between the remaining
photons, which is the property of tripartite GHZ state. Indeed, one
can easily show that the $|1,2\rangle$ state (\ref{eq:GHZstate})
always reduces to a product state, if one photon is not measured.
The reason for this is that if one photon is traced away, then the
remaining photons is put into a definite mode because of our
assumption of perfect phase matching and the resulting state is a
mixed state of the form
\begin{equation}
\rho=\sum_{\vec{k}} |F(\vec{k})|^2 |\vec{k}_{p}-\vec{k}, \vec{k}\rangle \langle \vec{k}_{p}-\vec{k}, \vec{k}|.
\end{equation}

Recently, we have found that to some extent, the $|1,1,1\rangle$
state can mimic some properties of the $|1,2\rangle$ state, e.g.,
by sending two nearly degenerate photons in the $|1,1,1\rangle$
state to the object while propagating the third one through the
imaging lens in the quantum imaging configuration, a factor-of-2
spatial resolution enhancement is achievable in the coincidence
counting measurement. However, the Gaussian lens equation is not
the same as that with the $|1,2\rangle$ state and more
importantly, the physics behind these two imaging processes is
quite different.

\section{Conclusion}
In summary, we have shown that the triphoton $|1,1,1\rangle$ state
is analogous to a W state, while the $|1,2\rangle$ state is
analogus to a GHZ state by comparing the third-order and
second-order correlation functions in both temporal and spatial
domains. Our analysis on these state properties may be important
to not only the understanding of multipartite systems but also the
technologies based on quantum entanglement. For example, in
Refs.~\cite{wen1} and \cite{wen2} we have discussed quantum
imaging using these two classes of states and have found different
spatial resolutions in application. The essential difference
between these two states is that $|1,1,1\rangle$ has a larger
Hilbert spaces than $|1,2\rangle$. Specifically, measurement of
one of the degenerate photons in the GHZ-type state allows for the
possibility of a separate measurement of the degenerate photon
state. This reduces the two-photon state to a mixed state.  For
the W-like state, only partial information can in principle be
obtained and so some entanglement remains.

The authors wish to thank their colleague Kevin McCann for help with
the numerical computations for the figure. We thank one of the
referee's for pointing out reference \cite{njp} to us.  We
acknowledge the financial support in part by U.S. ARO MURI Grant
W911NF-05-1-0197.

\end{document}